\documentclass{sig-alternate}

\usepackage{algorithm}
\usepackage{algorithmic}
\usepackage{subfigure}
\usepackage{mathptmx}
\usepackage{courier}
\usepackage{amsmath}
\usepackage{amsfonts,amssymb}
\usepackage{graphicx}

\newcommand\T{\rule{0pt}{2.6ex}}
\newcommand\B{\rule[-1.2ex]{0pt}{0pt}}

\begin{document}

\title{Personal Marks and Community\\
Certificates: Detecting Clones in Mobile\\
Wireless Networks of Smart-Phones}

\numberofauthors{2}

\author{
\alignauthor
Marco Valerio Barbera\\
       \affaddr{Department of Computer Science}\\
       \affaddr{Sapienza University of Rome}
\alignauthor
Alessandro Mei\thanks{This work was performed while Alessandro Mei
was a Marie Curie Fellow at the Department of Computer Science and
Engineering, University of California San Diego, USA. The fellowship
is funded by the European Union Seventh Framework Programme (FP7/2007-2013)
under grant agreement n.\ 253461.}\\
       \affaddr{Dept of Computer Science}\\
       \affaddr{Sapienza University of Rome, and}\\
       \affaddr{Dept of Computer Science and Engineering}\\
       \affaddr{University of California San Diego, USA}\\
 }

\maketitle
\begin{abstract}
We consider the problem of detecting clones in wireless mobile ad-hoc
networks. We assume that one of the devices of the network has been
cloned. Everything, including certificates and secret keys. This can
happen quite easily, because of a virus that immediately after sending
all the content of the infected device to the adversary destroys
itself, or just because the owner has left his device unattended for a
few minutes in a hostile environment. The problem is to detect this
attack. We propose a solution in networks of mobile devices carried by
individuals. These networks are composed by nodes that have the
capability of using short-range communication technology like
blue-tooth or Wi-Fi, where nodes are carried by mobile users, and
where links appear and disappear according to the social relationships
between the users. Our idea is to use social physical contacts, securely
collected by wireless personal
smart-phones, as a biometric way to authenticate the legitimate
owner of the device and detect the clone attack. 
We introduce two mechanisms: Personal Marks and Community Certificates.
Personal Marks
is a simple cryptographic protocol that works very well when the
adversary is an insider, a malicious node in the network that is part,
or not very far, from the social community of the original device that
has been cloned. Community Certificates work very well when the
adversary is an outsider, a node that has the goal of using the stolen
credentials when interacting with other nodes that are far in the
social network from the original device. When combined, these
mechanisms provide an excellent protection against this very strong
attack. We prove our ideas and solutions with extensive simulations
in a real world scenario---with mobility traces collected in a real
life experiment.
\end{abstract}




\section{Introduction}

You have left your
smartphone on the table at a cafeteria. You soon realize and go back to take it. Fortunately,
it is still there! You feel safe---while you are not safe at all. An adversary has connected your
smartphone to a laptop and dumped all of its memory, including public and secret cryptographic
keys. It is a matter of seconds or, at most, minutes.
You don't revoke your certificates (you feel safe!) and, a month later, you discover
that your credentials have been used by someone else. If you think this cannot happen---people
take very good care of their personal devices---consider that according
to a fairly recent report (WTOP, 15 Nov 2006) 478 laptops have been lost or stolen from the IRS (the Internal
Revenue Service is the United States federal government agency that collects taxes and enforces the internal revenue laws) between 2002--2006; 112 held sensitive taxpayer data, including SSNs.

Portable personal devices---smart-phones, laptops, and PDAs---are more and more
used in our everyday life. We use them to make phone calls, to plan our activities, to surf
the web, to manage our banking account, and probably very soon to make purchases.
The system we consider is a network of personal smart-phones that connects to
each other by using short-range communication technology like blue-tooth.
The nodes are the devices and the links appear and disappear as
people move and get in physical touch. As a consequence, the network is not a
collection of randomly moving objects---it has a social structure that can be exploited to
deliver revolutionary applications, and, we believe, new and surprisingly effective solutions to classical
networking and security problems.

These networks have been called with several different names---\emph{pocket switched
networks}~\cite{hui05, hui06}, \emph{mobile social networks}, \emph{opportunistic mobile ad-hoc
networks}, among others---and have drawn the attention of many researchers in the community.
Most of the work has focused on forwarding, that is how to route messages in such
a way to deliver them as quickly as possible and as cheaply as possible to
destination. Here, we focus on security. The attack that we have described
in the first rows of this paper is the \emph{clone attack}. The clone attack
might as well be performed by a virus that infects our device and sends all the
data, including secret keys associated with the certificates, to the adversary.
So, you are not safe even though you never forget your
smart-phone at the coffee shop. In this paper we are looking for a solution to this problem.
Of course, we do not want one
that consists in typing a password every time we use our certificates. People don't like
passwords and tend to forget them, or, even worse, choose trivial ones.

Our idea is to use social physical contacts, securely collected by wireless personal
smart-phones, as a biometric way to authenticate the owner of the device.
Indeed, our social physical contacts---our family members, the same barman everyday at the coffee shop,
our colleagues at work, our old friends when we hang out and relax after
work---characterize ourselves in a distinctive way.
Of course, each day is somewhat different and we don't always
meet the same people, but surely
there is a strong regularity in the communities we live in and in
circle of friends that we usually meet in person.
We will see how to use this regularity, complemented with other essential
mechanisms, to detect the clone attack in wireless networks of smart-phones
or other personal devices and to prevent the misuse of stolen certificates.

In this paper we introduce two protocols: Personal Marks and Community Certificates.
Personal Marks is a simple cryptographic protocol that proves to be an excellent way to
detect the attack when the adversary is an \emph{insider}, a person that belongs to the same
community of the victim. Community Certificate is a solution based on certificates
that tells how the node is expected to behave in terms of social contacts. If the clone
is an \emph{outsider}, a node that behaves in a different way with respect to the victim and meets different nodes and
different communities, then Community Certificate works very well,
the certificate soon expires, and the clone cannot
authenticate any more to the other nodes of the network. The two protocols are meant to be
used at the same time and, according to a large set of experiments made with
well-known traces of human contacts computed during real life experiments, they
collectively prove to protect the nodes against the clone attack in an excellent way.

\section{Related Work}

The detection of the clone attack is one of the most investigated security issue in wireless ad-hoc networks. 
As far as static wireless networks are concerned, there are three main approaches to the problem: Centralized, local 
and distributed random based techniques. The centralized techniques~\cite{egCCS02, bgpvk07} require nodes to send 
either location information or key information along with the node ID to a base station which in turn processes 
the information received and, in case of anomalies (same node ID with different locations), trigger a revocation 
procedure. Aside from presenting a single point of failure (the base station), this approach has the drawback
of incurring high communication cost to the network due to the high number of messages involved.

Local-based schemes~\cite{cps03, d02, egCCS02, nesp04} make use of voting mechanisms within nodes' neighborhoods to 
detect clones. Though they don't suffer from the drawbacks of the central based schemes, they fail on detecting  
replication attacks where node clones are scattered in different areas of the network.
 
The distributed and random based techniques~\cite{ppg05, cdpmm07, cdpmm10, zsjrw10} require nodes to send signed location information to randomly selected destinations on the network in a hop-by-hop fashion. These schemes rely on the high probability of intersection
of different location-declaration routing paths started from replicas of the same node. The nodes on which such intersection occurs are called witnesses, and their task is to then trigger node revocation procedures.
Another distributed scheme~\cite{czlp07} uses a random value, distributed to the nodes by the base station, to generate independent clusters in the network. Each cluster is an Exclusive Subset Maximal Independent Set (ESMIS) 
and has a cluster head denoted with SLDRs. One or more trees (whose nodes correspond to SLDRs) are defined over the network graph, and bottom-up aggregation protocols on such trees are used to detect node replicas (that are present in more than one ESMIS).

Although these distributed techniques do not present single points of failure, the overhead they incur either in message traffic or in computational terms~\cite{czlp07} is far from negligible. Moreover, all the aforementioned techniques, by relying on fixed geographical 
position of the nodes in the network are not apt to be used in mobile scenarios such the one we consider~\cite{hui05, hui06}.

With the increasing use of cellular technology, cloning of mobile devices for the purpose of making fraudulent telephone calls become a 
real threat to phone carriers. While this was a real threat in the nineties when CDMA was adopted~\cite{wikipedia}, in the GSM era it 
started to be considered as a minor/non feasible attack, at least for fraudulent billing purposes. 
With the spreading of the mobile pay 
systems (72.8 million of users at the end of 2009, expected 220 million by 2011 in China only~\cite{mobPayChina}), and especially those 
based on credit card transactions, phone cloning has become a vicious threat to the users portfolio~\cite{sbk07}. Thus a lot of 
research on detecting such attacks have been done, most of which is based on the use of neural networks
by the carrier to detect possible anomalies. For a good survey see~\cite{survey}.

The idea of exploiting information regarding social ties between nodes is not new, actually it is common to a good part of the
literature on pocket switched networks (PSN) and similar social networks. Much research has been dedicated to the analysis of the data
collected during real-life experiments, to compute statistical properties of human mobility, and to uncover
its structure in sub-communities~\cite{toronto, hui05, hui06, milan07, UCAM-CL-TR-617, cai07mobicom, cai08mobihoc}.
Later on, most of the work in the field focused on message forwarding and to find the best strategy to relay messages
in order to route them to destination as fast as possible (see~\cite{simbet,hui08mobihoc,dfw08}, among many others).
Also security problematics such as node capture~\cite{cdpgma10} or selfishness~\cite{routingSelfishBuffer,g2g} have been solved by
making use of social relationships among nodes. 

Biometrics have been used for centuries to authenticate people. The most common example of biometrics is
the handwritten signature that we use everyday to certify our identity. With technological development,
the use of biometrics to authenticate has become a well-known area of investigation in computer security.
Several biological measurements have been used to identify people to computer systems: voice, face,
iris, keystroke dynamics (the way we type is often sufficiently unique to identify ourselves), and others.
In many cases these are not fully automatic. With keystroke dynamics, for example, the user has to type.
As an excellent reference and starting point on computer biometrics, see~\cite{biometrics}.

The authors in~\cite{Barbeau05} propose two intrusion detection systems that have similar biometrics
ideas: The first one is build upon Radio Frequency 
Fingerprinting (RFF), whereas the second one leverages User Mobility Profiles (UMP).
However both detection systems are centralized, and rely on the fact that the intruder (the clone) behaves substantially
differently from the real user in terms of geographical movements.
Thus, compared with the solutions proposed in this paper, both systems are based on a completely different
idea and are not able to detect anomalies when the clone behaves similarly to the original node (for example,
when the clone attack happens in a building). Lastly, the solutions are not distributed.

To the best of our knowledge, this is the first paper that presents biometric authentication techniques based on the
social contacts of the owner or the device to detect the clone attack in mobile wireless networks of smart-phones.
The techniques, namely \emph{Community  Certificates} and \emph{Personal Marks}, are \emph{only} build upon the
social-guided meeting patterns of users
in the mobile social network. They are thoroughly orthogonal to trust or voting mechanisms, or geography based techniques to detect 
cellphone theft or credit card fraud, that are based on different ideas and can be used at the same time.

\section{The System}
\label{sec:systemmodel}

Our network setting is made of last generation smart-phones. These devices are more
and more popular---we can easily envision a society in which almost everybody will carry
a personal computing device of this kind. Commercial smart-phones are usually able to communicate
by using short-range communication technology, like blue-tooth and/or Wi-Fi, aside classic GSM.
When two people meet, their personal devices establish
a link by using one of the above mentioned short-range technologies.
In the network we consider, nodes are devices carried by people, and links appear and
disappear as people move and meet.

Smart-phones are not-so-small
devices that can easily handle video/audio streaming, 3D games, web surfing and SSL sessions,
and other applications. Therefore, we can safely assume that nodes are able to perform
public key cryptography that is used to sign messages and to establish secure communication sessions
among peers.  The nodes are equipped with public/private key pairs, and
the former is signed by a trusted authority CA.
We assume
that our protocols execute by setting up authenticated and encrypted sessions among
the peers. 
Nodes are loosely time synchronized. Loose time synchronization is very easy to
get, if a precision in the order of the second is enough, like in our protocols.
We also assume that the trusted authority is able to send a message
to any node in the system, for example using the cellular network. When a clone is 
present, the message is received by the original node and by the clone as well
(of course it is perfectly possible that the clone has turned off its interface to the cellular network).

Our protocols rely on authenticated logs of physical contacts between devices. However,
if not carefully implemented, these logs can be attacked. We assume that the nodes use
techniques like distance bounding~\cite{distancebounding,sector}
to guarantee that the log of a contact between two devices can be collected only when
the devices are provably in physical proximity. These mechanisms are based on measuring the delay in the
communication.

Lastly, we assume that the users have access to an alternative way
to authenticate to the authority. 
There are several examples of such mechanisms. One example is GMail: If you forget
your password, you can still authenticate by responding to a list of personal questions
that, most probably, only you can respond. In other systems, you might be able to
authenticate by using a smart-card at your desktop at home. Another example is the
pair of PIN and PUK codes used in cellular phones, the first one is short and easy
to type, the second one longer, more tedious,
and is meant to be used only to recover from PIN blocking
or other uncommon events.
In any case, we assume that
the alternative mechanism to authenticate is secure but long, burdensome, and
we definitely want to use it only in rare and exceptional
circumstances like when we need to recover from a clone
attack.

In the rest of this paper, we use $\langle m\rangle_i$ to denote a message~$m$ signed by node~$i$.

\subsection{The Adversary and the Problem}

We consider the following scenario: A smart-phone is infected by a worm
or by a virus. The goal of the virus is to make a perfect copy of the device of the victim, to send all
the data to the adversary's device, and to destroy itself without leaving any trace. A similar attack
can be performed in a different way: You left your device unattended for just a couple of
minutes; when you go back to take it, it is still there. What you don't know is that
the adversary has made a clone by copying all the memory of your device.

Later on, the clone can be used to interact in the network by using the victim's credentials.
Note that everything has been cloned, including certificates and secret keys. We
assume that the clone can be switched off arbitrarily by the adversary to prevent detection---the
device might be turned on only when necessary (that is, when it is used to connect to the other peers in the network with
the stolen credentials). To start with, we assume that the adversary has been able to clone only
one device. Later in the paper we will discuss what is going to happen if more devices
can be stolen.

Our goal is to build a system able to detect and/or stop as fast as possible and as accurately
as possible this attack.

\section{Detection of the Clone Attack} 
\label{sec:solution}

The users of our system are humans---their mobility pattern, the people they meet, and the places 
they visit are not random. Rather, the dynamics of the network is highly influenced by the
fact that almost everything in our everyday life is guided by social relationships
(e.g.\ friends, family) and interests (e.g.\ work, school, gym, or playing chess).
If we think of our daytime, it usually follows
a  do-repeat cycle that might look like this: Go to work (or school); meet with colleagues; finish work and go out with
the same old friends
or do whatever we like to do during our free time; go back home; go to work again and so on. Of course,
it is not always like that (fortunately), but surely there is a regularity in the things we do and the people we meet.
The same cycle is performed by our mobile device, and the ones of our friends, family members, or colleagues
that we meet over and over again. The people (devices) we meet
everyday characterize ourselves in a remarkably distinctive way.
One of the ideas in this paper is to use this characterization
as efficiently as possible in order to make nodes prove who they are not with something they know (passwords),
or something they have (certificate), but with the proof that they actually meet physically and regularly the people
they are supposed to meet.

\subsection{Personal Marks}
\label{sec:marks}

We start by introducing Personal Marks. Personal Marks is a very simple cryptographic protocol we can use
to distinguish clones. Actually, Personal Marks is able to tell you that the person that you have before
is not the same person that you met last time, though it can't really tell who is the impostor.

Let's make a step back and proceed more formally.
Personal Marks consists in \emph{marking} each node in your community with a small
cryptographic object, signed by yourself, that you can check to have some level of guarantee
that node~$j$ is exactly the same node~$j$ you have met earlier. First of all, we call
\emph{community} of node~$i$ its set of friends---the set of other nodes that gets frequently
in physical contact with node~$i$. One way to compute the community of node~$i$ is by using
the distributed community detection 
algorithm described in~\cite{hui07}. Communities are a fundamental notion in social networks,
people meet more often other people of the same communities and this intuitive property 
have been leveraged in the literature in several ways.

Personal Marks works as follows: When two nodes of the same community~$i$ and~$j$ meet, they both
put a mark on the other device. The mark that node~$i$ gives to node~$j$ contains a timestamp
signed by node~$i$ itself. Same is done by node~$j$ and given to node~$i$. The next time they meet,
they both request the mark to check it before starting the new session. 

If there is an clone present in the community, say a clone~$j'$ of original node~$j$, and~$j'$ gets
in touch with the community of~$j$, Personal Marks is going to detect the anomaly.

Assume that node~$j$ has been cloned at time~$t_0$ and that node~$i$ gets in physical
contact with node~$j$ for the first time after the attack at time~$t_1>t_0$.
Clearly, node~$i$ does not know if the node is the original or the clone and, by using
Personal Marks, it freshens the mark with the node. The detection is triggered when
node~$i$ meets the \emph{other} node, that cannot help but exchanging the old
version of the mark, a mark with a timestamp in the interval~$(-\infty,t_0)$ that does
not pass the test by node~$i$. If the adversary is an insider,
node~$i$ will soon meet both the nodes, the clone and the original one (that is a friend).
While node~$i$ is not able to tell who is the original node and
who is the impostor, the
clone attack is detected and node~$j$ is revoked by the authority.
The legitimate node~$j$ has now to go through an alternative, longer procedure to re-authenticate.

\subsubsection{Personal Marks: The Check and the Update Protocols}

In our system, a personal mark is a small cryptographic object that carries some randomness from
both peers. The authentication protocol in Personal Marks is made of two parts:
The mark-check, and the mark-exchange.
Let us start from the mark-exchange protocol. Each time two nodes~$i$ and~$j$ meet, they contribute
some randomness~$r_i$ and~$r_j$ to freshen the mark (or to establish one if this is the first
time the nodes meet). The mark is of the form
$\langle\text{MARK},t,r_i,r_j\rangle$, where $t$ is a timestamp (the timestamp is not crucial
for the system to work, but it gives information about the timing of the attack in case of
detection). The mark is signed and stored by both peers. Note that, if the protocol is not
completed (may be for lack of continuos connectivity), the peers can safely assume that
it has not happened.
\begin{figure}
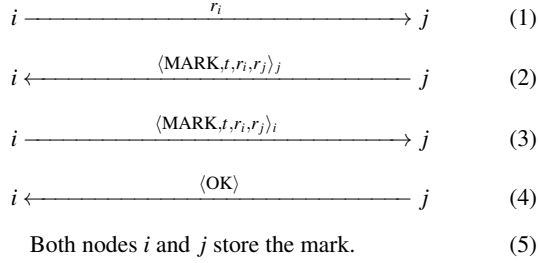

\begin{align}
i & \xrightarrow[\hspace*{5cm}]{r_i} j\\
i & \xleftarrow[\hspace*{5cm}]{\langle \text{MARK},t,r_i,r_j\rangle_j} j\\
i & \xrightarrow[\hspace*{5cm}]{\langle \text{MARK},t,r_i,r_j\rangle_i} j\\
i & \xleftarrow[\hspace*{5cm}]{\langle \text{OK}\rangle} j\\
  & \text{\; Both nodes $i$ and $j$ store the mark.}
\end{align}
\caption{Mark update protocol. Here we assume $ID_i<ID_j$.}
\label{fig:markUpdate}
\end{figure}

Let us now look at the mark-check protocol, executed before the mark-exchange
protocol when it is not the first time that nodes~$i$ and~$j$ meet.
Suppose, without loss of generality, that first node~$i$ checks node~$j$ and then node~$j$
checks node~$i$. Node~$i$ sends a request to which $j$ has to reply with the last version of the mark.
After receiving the mark from~$j$, $i$ checks the signature and the timestamp and
notifies node~$j$. In the case when the test is failed, both the correct mark and the mark exchanged during this protocol
can be sent to the authority by using GSM or broadcast in the network as a proof of the clone attack, and the credentials
of node~$j$ are thus revoked. Of course, the legitimate node~$j$ is also invited to re-authenticate and get new
credentials.
\begin{figure}
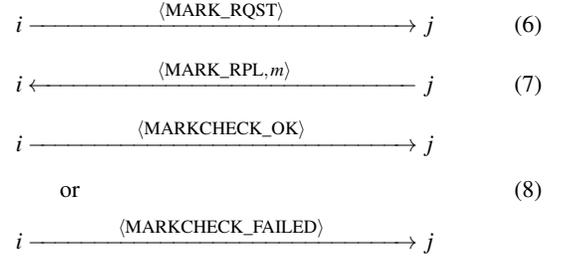

\begin{align}
i & \xrightarrow[\hspace*{5cm}]{\langle \text{MARK\_RQST}\rangle} j\\
i & \xleftarrow[\hspace*{5cm}]{\langle \text{MARK\_RPL},\,m\rangle} j\\
\nonumber i & \xrightarrow[\hspace*{5cm}]{\langle \text{MARKCHECK\_OK}\rangle} j\\
& {\hspace{5mm}\normalsize\text{or}}\\
\nonumber i & \xrightarrow[\hspace*{5cm}]{\langle \text{MARKCHECK\_FAILED}\rangle} j
\end{align}
\caption{Mark check protocol. Here $CA$ denotes the trusted authority.}
\label{fig:markCheck}
\end{figure}

Note that if a clone refuses to initiate or to complete either of the two protocols with the goal of
avoiding detection, then it does not authenticate and its cloned certificates are going to expire soon
thanks to Community Certificates,
the other mechanism we propose in this paper that works very well when the adversary does not meet the
members of the communities of the original owner.

\subsubsection{Personal Marks, Insiders, and Outsiders}

Personal Marks works on the assumption that both the clone and the original node get in touch
with a common node in the community of the original node, sooner or later. This is often true---the
adversary is someone that is part of the same community of the victim: A colleague
at workplace, or a fellow student at college. In this case, we will say that the adversary
is an \emph{insider}, and Personal Marks works extremely well as shown later in the
experiments with real mobility traces.

However, this is not always the case. It is perfectly possible that the clone uses the credentials
that it has stolen far from the communities of the original node in the network.
In this case, we will call the adversary an \emph{outsider}. This is also a very important scenario, in many real
cases the attack can be performed by someone that we don't know and that has the
goal of taking our device to a different city or in a different part of the same city
and use the credentials when needed.
Clearly, Personal Marks can't do much against outsiders. Outsiders never meet someone
from the community of the victim. Therefore, they are never checked and they don't need any
fresh mark. To fight back outsiders, we have to develop a completely new mechanism.

\subsection{Community Certificates}
\label{sec:communityCertificates}

When a node, say node $i$, agrees to enter the system, it automatically enters a training period during which it securely collects 
signed and timestamped logs of the physical contacts with other nodes. At the end of the period the logs are reported to the
authority. The authority uses the logs to build a signed certificate $ComC_i$ that is sent back to node~$i$. Clearly, all these
messages are encrypted and authenticated, like all sessions in our protocols. The certificate is of the form
$ComC_i = \langle\langle \mathit{FS}_i, \mathit{FI}_i, k_i\rangle_{CA}, \mathit{SU}_i\rangle$. In the certificate,
$\mathit{FS}_i$ is the set of ``best friends'' of node~$i$ (its community);
$\mathit{FI}_i$ is a mapping that tells the ``strength of the relationship'' between
node~$i$ and his best friends. More specifically, for every $j\in \mathit{FS}_i$, $\mathit{FI}_i(j)$ is a
value computed as a monotone function of the inter-contact times between nodes~$i$ and~$j$ observed during the training period.
Moreover, for every $j\in \mathit{FS}_i$, $\mathit{SU}_i(j)$ is a timestamp signed by node~$j$ that certifies
at what time there has been the last contact between node~$i$ and~$j$ (this object is similar to the
mark in Personal Marks). Given node~$j\in\mathit{FS}_i$, we say that the timestamp~$\mathit{SU}_i(j)$
is \emph{fresh} if $t<\mathit{SU}_i(j)+\mathit{FI}_i(j)$.
Certificate~$ComC_i = \langle\langle \mathit{FS}_i, \mathit{FI}_i, k\rangle_{CA}, \mathit{SU}_i\rangle$
is valid at time~$t$ if and only if at least $k$ signed timestamps in set~$\mathit{SU}_i$ are fresh.
In other words, the certificate is valid only if the node has been able to collect enough fresh
signatures by physically meeting people in his circle of friends.
When the authority generates the certificate, $\mathit{SU}_i$ can be prepared with timestamps
signed by the authority, just to make the certificate immediately valid.

Note that all this process is totally automatic. Note as well that we assume that no attack occurs during the training period.
This is quite reasonable indeed, if such an attack takes place, the authority would receive training logs from both the victim
and the clone at the same time, thus revealing the attack.
 
\subsubsection{Community Certificates and Outsiders}
\label{subsection:commcert}

The community certificate is a requirement for the authentication of node~$i$ to the other peers of the system.
Indeed, when $i$ meets another node, say node~$j$, it is required to show the certificate $ComC_i$
in case it is willing to set up a communication or to use one of the services provided by node~$j$.
The simple protocol is described in Figure~\ref{fig:authentication}.
\begin{figure}
\begin{align}
i & \xleftarrow[\hspace*{5cm}]{\langle \text{AUTH\_RQST}\rangle} E\\
i & \xrightarrow[\hspace*{5cm}]{\langle \text{AUTH\_RPL},\,ComC_i\rangle} E\\
\nonumber i & \xleftarrow[\hspace*{5cm}]{\langle \text{AUTH\_OK}\rangle} E\\
& {\normalsize\text{or}}\\
\nonumber i & \xleftarrow[\hspace*{5cm}]{\langle \text{AUTH\_DENIED}\rangle} E
\end{align}
\caption{Community-Authentication protocol: Node $i$ authenticates to entity $E$
(either a node or a resource like an ATM machine/card reader etc.).}
\label{fig:authentication}
\end{figure}

As mentioned, an adversary that has the goal of using the cloned smart-phone outside the communities
of the legitimate owner is an outsider. Community Certificates proves to be an excellent mechanism to
stop clone attacks by outsiders. Indeed, when taken outside the communities of the owner, the certificate
expires soon. The adversary is simply not able to collect enough fresh signatures to keep the
certificate valid.

In the experiments with real traces, we show that it is possible to choose $\mathit{FS}_i$, $\mathit{FI}_i$,
and $k$ in such a way that the certificates are continuously and consistently valid, when carried by
the legitimate owner, and expire quickly when carried by outsiders.
Therefore, a combined use of Community
Certificates and Personal Marks is an excellent way to provide efficient and secure authentication and to thwart the
clone attack in mobile wireless networks of smart-phones. In Section~\ref{sec:caveats} we discuss
a few ideas, add-ons, and solutions to extend the features of Community Certificates and to deal with
non-frequent scenarios like the case when the legitimate
owner has to use his own community certificate during a long travel far from his usual communities.
 
\subsubsection{Dynamic Community Certificates}

So far, we have described Community Certificates as a static system. However, in real life it may be possible, even
if it is not common, that we change our own community. In general, it is reasonable to imagine the following scenarios:
(i) our community changes completely since, for example, we move to another town;
and (ii) one of our friends moves away, or a new node is our new best friend. Here, we
see that it is easy to design protocols to dynamically change the community certificate in a secure
way.

In case (i), it is enough to start off a new training phase and to get a new certificate. In case (ii),
we can initiate a selective update of the certificate to remove one node, or to add a new one,
or to update the parameter of a node that is already part of our ring.
Of course, the addition and/or the removal can change all the parameters of the certificate, like
mapping $\mathit{FI}_i$ or $k_i$. The procedure can be easily secured. Indeed, when the
procedure starts, the authority sends a GSM message to the node. If a clone requests
the procedure to change the certificate according to his own communities, than the
message is received by the original owner as well, that promptly detects the attack
and sends to the authority a signed request of certificate revocation.

\subsection{Multiple, Coordinated Clone Attacks}
\label{subsection:multiple_attacks}

In this paper, we consider the problem of detecting a single clone attack. Generally speaking, if the adversary
is very powerful, it is however possible that it clones a whole set of mobile nodes. While we don't explicitly
deal with this case in this work, it is still useful to see what is going to happen with Personal Marks and
Community Certificates.

What we can say is that Personal Marks keeps working with (approximately) the same performance.
According to our experiments, communities are not closed sets of nodes. Usually nodes belong to many different
communities and the structure of the network is characterized by many \emph{overlapping} communities.
This is in accordance with other well-known research work in the literature~\cite{palla05}. In other words,
it is usually impossible to select a small community of nodes that have very little contacts with nodes outside
the community (of course, if you can select the whole network than it is easy, but if the adversary is able to clone
the whole network that there is little we can do about it).
Therefore, Personal Marks is still efficient in detecting the attack, thanks to all the
interactions between the nodes of the cloned community with nodes outside the community.

A similar argument is true for Community Certificates---it is usually impossible to isolate a small community
whose nodes do not have nodes that are outside the community in the certificate. However, it is in principle
possible to isolate and clone a community in such a way that the cloned devices can authenticate each other.
To make this attack possible only if the community to be cloned is very big, parameter~$k$ must be
set high enough, and certificates should be computed in such a way not to form cliques. While this is
out of the scope of this paper, it is certainly a open and interesting research direction. In any case,
that means that the adversary is able to hijack a whole community. Moreover, we will see how to
extend Community Certificate by admitting fixed devices in the certificate. For example, a fixed
infrastructure in both our home and our office. While cloning a device can be easy, stealing a fixed
infrastructure can be considerably harder.

\section{Experimental Results}
\label{sec:experiments}

In this section, we present some experimental results in order to show the detection performances of our system. The results refer to the two typologies 
of attack we consider: Insiders and outsiders. In the first case the detection relies on the Personal Marks sub-system. Rather, in the second, the 
Community Certificates sub-system comes into play and the detection is determined by the expiring time of the certificate when in the hand of the adversary.  We start off by giving detailed information on the dataset we used to evaluate our protocol. Then we present the results obtained with
Personal Marks and with Community Certificates.

\begin{table*}[t]
\begin{small}
    \caption{Details on the datasets (DS) and respective training period (TR).}
    \label{tab:traces}
    \begin{center}
    \footnotesize{
        \begin{tabular}{|l|c|c|c|c|c|c|}\hline
            Data set \T \B & Dartmouth & UCSD & Reality & SWIM\\ \hline
            Total nodes	& 1101 & 32 & 45 &  1500\\
            DS AVG active/day & 1034 & 27 & 37 &  1500\\
            TR AVG active/day & 980 & 28 & 38 &  1500\\
            DS AVG contacts/node/day & 283 & 49 & 15 &  132\\
            TR AVG contacts/node/day & 263 & 51 & 16 &  131\\
	DS AVG stability/node/week & 0.55 & 0.52 & 0.55 & 0.49\\ [1mm] \hline
        \end{tabular}}
    \end{center}
\end{small}
\end{table*}

\subsection{Datasets}
\label{subsection:datasets}

All the experiments have been performed by using a simulator driven by three type of traces:
WLAN (real), bluetooth (real), and simulated traces.
For the WLAN traces we used the \emph{Dartmouth}~\cite{dartmouth} and \emph{UCSD}~\cite{wtd} datasets. The bluetooth trace we used is the 
\emph{Reality}~\cite{mit-reality-2005-07-01}, collected during the Reality Mining project at the MIT Media Lab. Finally, the simulated trace has
ben generated using the \emph{SWIM}~\cite{mei:swim} model.

In contrast with the simulated scenario, in which we have been able to select the number of nodes and the length of the experiment, in the case
of WLAN and bluetooth traces we had to select a period of time reasonably long that contained a set of nodes that are fairly active.
The reason is that real traces often suffer from data loss and contain nodes whose activity is recorded only for a portion of the total
time span covered  by the trace. This might cause loss of performance of our system. 

\subsubsection{Dartmouth}

The Dartmouth trace consists of the SMNP logs of the access points across the Dartmouth College campus from April 2001 to June 2004.
To infer the contacts between the nodes we follow the assumption widely used in the literature that two nodes are able to communicate to
each other whenever they are associated to the same access point (\cite{ucsdpaper} \cite{hui06}). From this trace we have selected a time span of 8 weeks,
from January 5, 2004 to March 1, 2004, during which a set of $1101$ nodes have recored to have at least $50$ contacts a day for at
least the $80\%$ of the days. This ensures us that these nodes stay active during the whole period we used to evaluate our protocol.

\subsubsection{UCSD}

The UCSD trace is part of the Wireless Topology Discover project (WTD) \cite{wtd}. The trace consists of the logs extracted from PDA
carried by a set of about 275 freshmen students of the University of California, in San Diego. The trace spans a period of $11$ weeks
between Semptember 22, 2002 and December 8, 2002, during which each PDA periodically recorded the signal strength of all the APs
in its range. To infer the contacts between the nodes we again used the assumption that two nodes can communicate as long as they are
in the range of a the same access point. As also reported in \cite{ucsdpaper}, we have found that the trace is characterized by a steady
decline of the user population that especially affects the last two weeks. For this reason we decided to restrict our tests to the first $8$ weeks
and to use only the set of $32$ nodes that recorded at least 1 contact a day for at least the $80\%$ of the days.

\subsubsection{Reality}

As opposed to the Dartmouth and UCSD traces that used WiFi radio logs, the Reality trace has been collected using short ranged bluetooth
radios. More in detail, the trace included the bluetooth records collected by 94 cellphones distributed to student and faculty on the MIT
campus during 9 months (from September 2004 to June 2005).  We have chosen this particular trace since it is one of the few existing traces
which encompasses a large number of nodes communicating through bluetooth technology for a long period of time. Like
other traces, however, many nodes recorded very few to no sightseeings for long periods of time. In order to keep the amount of nodes high
we thus restricted ourselves to a  period of 8 weeks going from October 18th to December 13th 2004 and discarded the nodes that didn't record
at least 1 contact a day for at least the $70\%$ of the days. This selection yielded a final set of $45$ nodes.

\subsubsection{SWIM}

This trace has been generated using the SWIM model (\cite{mei:swim}) that has been shown to simulate well human mobility in conference and
university campus environments. The SWIM simulator has also been shown \cite{swim-secon10} to be able to properly scale a reference scenario by
keeping the nodes density constant. Thanks to this we have been  able to replicate the statistical and social properties of the Cambridge
Campus bluetooth data set \cite{cambridge06} (that is only 11 days long) increasing both the number of nodes
and the time span by keeping the same dynamics of the original real trace in terms of average number of contacts per user:
The generated trace contains $1500$ nodes and is $8$ weeks long. Since the simulator keeps track of all the contacts between the
nodes, it hasn't been necessary to preprocess the trace and we've been able to use the whole set of nodes.

\subsection{Training Period}

As we already discussed in the section \ref{sec:communityCertificates}, before a node can use Community Certificates, it is required to go
trough a \emph{training period} during which it collects logs of contacts. At the same time, the Personal Marks system described in
Section~\ref{sec:marks} assumes that each node know its community. For this reason we decided to split all the traces in two parts: the former
used as a training period for all the nodes, the latter used to test the performances of our protocols. During the training period, we both collect
data about the contacts between the nodes needed by the community certificates and extract the communities of the nodes. For this, we use
the $k$-clique algorithm~\cite{palla05}, one of the most popular in the area of social mobile
networking (\cite{hui05, hui07, hui08mobihoc, swim-secon10}),
of which there is a distributed version (\cite{hui07}).

Since the movement of people is usually characterized by a high degree of regularity, we have seen that it is not necessary that
the training period be long: In every trace we used a training period consisting of only the initial $25\%$ of the total time span of the trace.
This is enough for our system to quickly detect both insiders and outsiders.
A further evidence that our intuition is good is given by the results shown in Table \ref{tab:traces}, where
it can be seen that the properties of the trace during the training period are very close to those of the rest of the trace.
In particular, the last row of the table shows what we have called the average \emph{stability} of the nodes. We defined the stability of a node~$x$
as a measure of the average similarity between the set of nodes~$x$ met during the training period and those it 
met during each week of the rest of the trace. More in detail, if $T$ is the set of nodes that $x$ met during the training period and
$S_i$ is the set of nodes that $x$ met during the $i$-th week of the trace, the stability of $x$ is computed as 
\begin{equation*}
\frac{1}{n} \sum_{i=1}^{n} \frac{|S_i \cap T|}{|T|}
\end{equation*}
where $n$ is the number of weeks of the trace (training period excluded). As shown in the table, the average stability of the nodes is always close or greater than $0.5$ meaning that, on
average, a node meets every week at least half of the nodes it met during the training period.

\subsection{Communities}

In this section we briefly give details about the communities found in the various traces. To use the $k$-cliques community detection algorithm
we first had to extract a social graph from the contacts happening in the training period of each trace. We put the edges in the graphs in such a way that they reflect the repetitive social behaviour of the nodes. Since all the not-simulated
traces we used have been collected in university campuses, we expect the social connectivity to be given by students that, for example, frequent the same classes, study in the same library or 
live in the same dormitory. Once the social graph is built, the $k$-cliques community detection algorithm is run over it in order to find groups of nodes socially close to each other. However, it is important to note that according to the experiments
our protocols are pretty robust with respect to different definitions of ``friend'' used in the literature, as long as it is true
that friends meet regularly and frequently.

\subsection{Personal Marks vs Insiders}

To test the performance of Personal Marks, we run a trace-driven simulation where we make each
pair of nodes of the same community check and exchange marks each time they meet.

In order to simulate the attack, for every node~$i$ we do the following: For every node~$j$ in the
community of node~$i$ we generate a clone attack of node~$j$ against node~$i$ (node~$j$ is the
insider) and measure the time required by Personal Marks to detect the attack.
Performing the experiment is rather straightforward. As we described in Section~\ref{sec:marks},
a node in the victim's community detects the attack in two cases: The first case is when the node first meets the adversary and then the victim,
the second case is when the node first meets the victim and then the adversary. In the first case, the mark of the victim node will be found
to be outdated, while, in the second case, it is the mark of the adversary that will be found to be outdated. Thus, we measure
the average time it takes for any of these two cases to happen after the attack happened.
\begin{figure}[t]
    \begin{center}
        \subfigure[Dartmouth]
        {
            \label{fig:marks-dartmouth}
            \includegraphics[width=.22\textwidth]{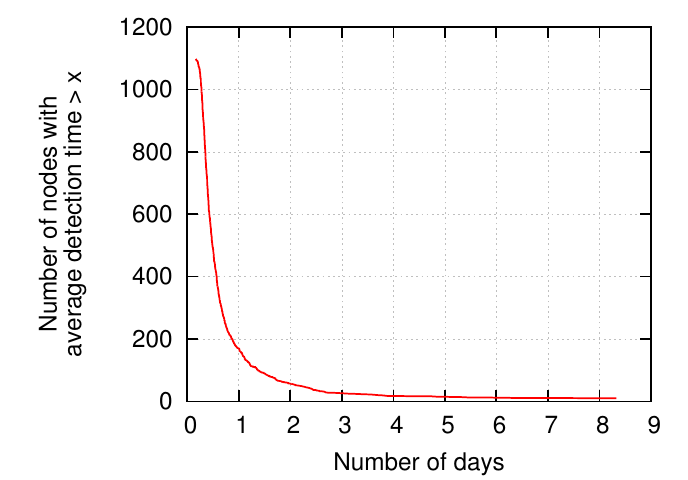}
        }
        \subfigure[UCSD]
        {
            \label{fig:marks-ucsd}
            \includegraphics[width=.22\textwidth]{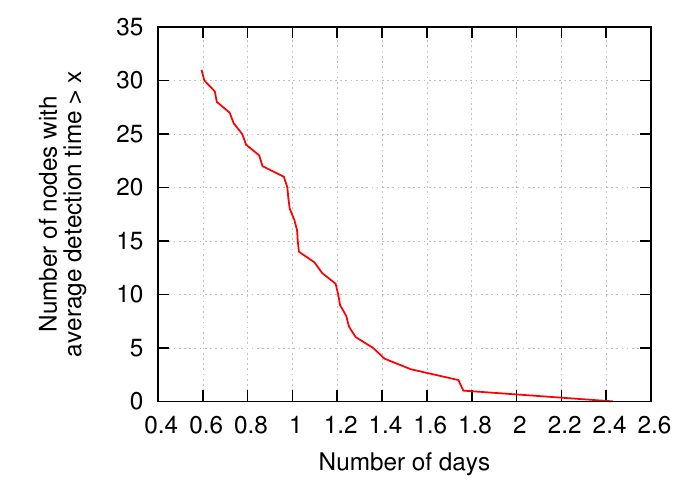}
        }
        \subfigure[Reality]
        {
            \label{fig:marks-reality}
            \includegraphics[width=.22\textwidth]{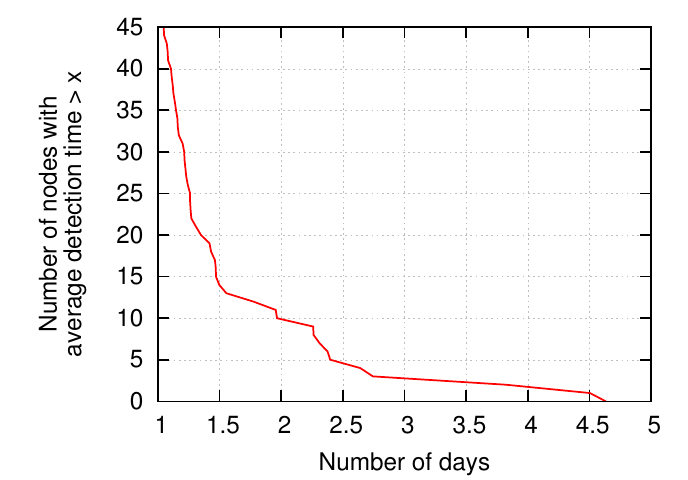}
        }
        \subfigure[SWIM]
        {
            \label{fig:marks-swim}
            \includegraphics[width=.22\textwidth]{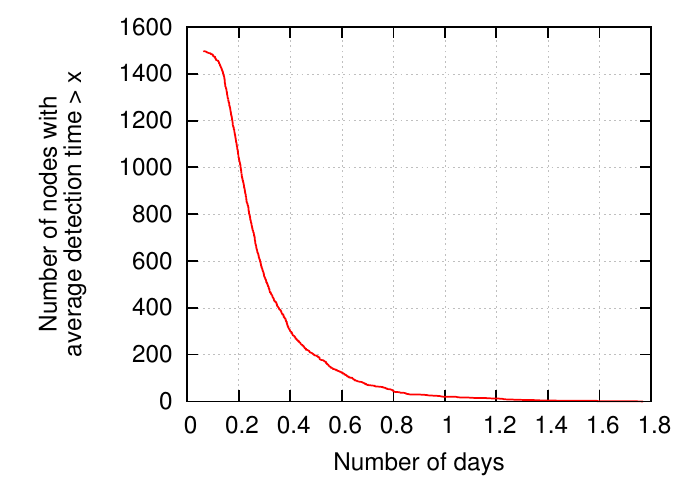}
        }\vspace{-2mm}
		\caption{Distribution of the average detection time of insiders}
        \label{fig:marks}
    \end{center}
\end{figure}

Figure \ref{fig:marks} shows the performance the Personal Marks sub-system on all the traces---real and synthetic ones.
We show, for a given number of days $x$, what is the number of nodes for which the detection of the attack takes
more than $x$ days on average.  First, we can notice from the results that, consistently in all the traces, the time needed to detect
an insider is very low. Indeed, in the majority of the cases the attack is detected in one day, one day and a half at most.
More in detail, in the Dartmouth trace (Figure~\ref{fig:marks-dartmouth}), 
the attack gets detected in less than one day in the $84\%$ of the cases and in less than two days in the $96\%$ of the cases.
In the UCSD trace the attack is detected in less than one day in the $44\%$ of the cases, but for the $97\%$ of the cases
it takes less than two days. In the Reality trace (Figure \ref{fig:marks-reality}) the
performance of Personal Marks is slightly worse: In the $69.5\%$ of the cases the attack is detected in less than one day and a half,
while it takes less than three days for the $93.4\%$ of the cases. The reason why it takes more time to detect the attack is that the Reality
trace is considerably sparser than the other traces. Indeed, Table~\ref{tab:traces} shows that the
average number of contacts is the minimum among all the traces. Lastly, Personal Marks shows the best performances in the SWIM
trace (Figure \ref{fig:marks-swim}): In the $86.5\%$ of the cases, the nodes need at most half a day to detect the attack, 
while in the $99\%$ of the cases it takes at most one day and a half. This is not surprising. In the real traces, in fact, all the nodes have
periods of time (that can last even a few days) where they didn't  report any contact. Of course, this is not very realistic---all
of the traces
capture only a very small fraction of the real contacts of the user (the other people in the experiment, often a
small subset of the other students at school). We expect that, in case of wide
deployment in the real world, the number of contacts would be much higher than any of these traces---both real and synthetic
ones---and the performance, already good, would greatly improved even compared to the synthetic traces, that
have been designed to mimic the dynamics of the Cambridge traces.

\subsection{Community Certificates vs Outsiders}
\label{subsec:commcert}

The Community Certificate sub-system has the goal to fight back outsiders.
This sub-system has three main components: Set $\mathit{FS}_i$ of nodes appearing in the certificate
of node~$i$, the values in $\mathit{FI}_i$,  and parameter $k_i$.

Let us concentrate on $\mathit{FS}_i$ and $\mathit{FI}_i$. The first one is the set of nodes from which node $i$ receives the signed
timestamps it uses to prove its identity. The second one defines for how long a signed timestamp given to $i$ by $j$ is \emph{fresh}
(see Section~\ref{sec:communityCertificates}). Clearly, the correct definition of $\mathit{FS}_i$ and $\mathit{FI}_i$ is crucial.
For instance, if the set $\mathit{FS}_i$ were large and the values $\mathit{FI}_i(j)$ for every $j$ in $\mathit{FS}_i$ were big, 
then it would be easy for node $i$ to prove its identity. In fact, the probability that node~$i$ meets at least $k_i$ nodes in the set~$\mathit{FS}_i$ would be large, and, at the same time, the signed timestamps $\mathit{SU_i(j)}$ that node $i$ would receive from them would last for long time.
This would surely translate into a low number of false positives in our system (legitimate users that cannot authenticate).
At the same time, however, this would give more chances to the adversary to keep his certificate valid for a long period of time.
Thus, by choosing the elements of set $\mathit{FS}_i$ and the values of $\mathit{FI}_i$ we trade-off the ability of the system to correctly
authenticate honest nodes and the performance in blocking outsiders.

\begin{figure}[t]
    \begin{center}
        \subfigure[Dartmouth]
        {
            \label{fig:fs_size-dartmouth}
            \includegraphics[width=.22\textwidth]{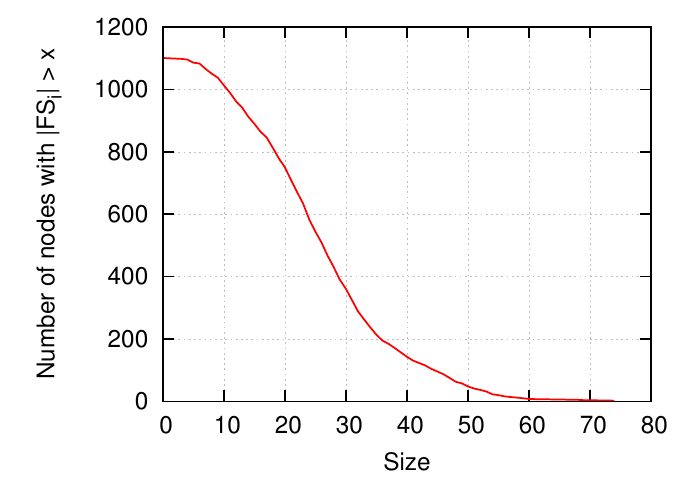}
        }
        \subfigure[UCSD]
        {
            \label{fig:fs_size-ucsd}
            \includegraphics[width=.22\textwidth]{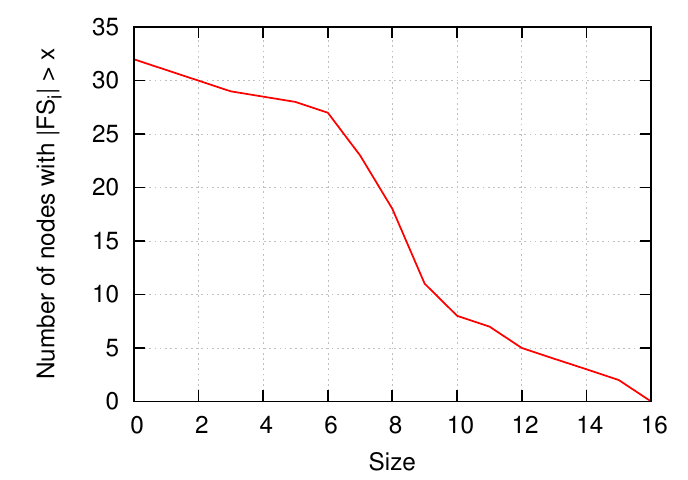}
        }
        \subfigure[Reality]
        {
            \label{fig:fs_size-reality}
            \includegraphics[width=.22\textwidth]{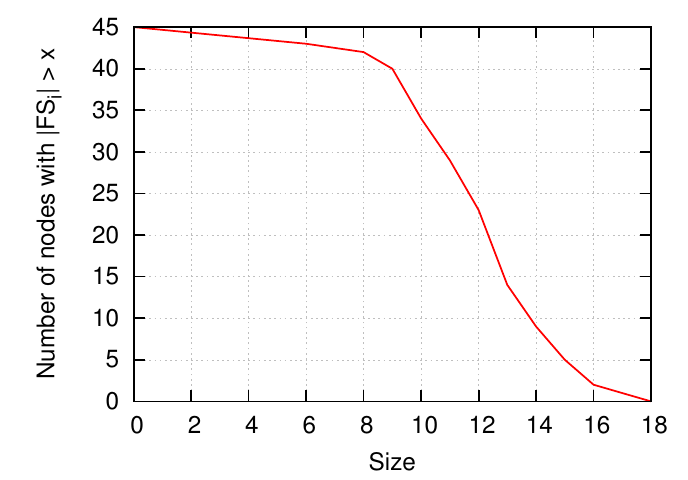}
        }
        \subfigure[SWIM]
        {
            \label{fig:fs_size-swim}
            \includegraphics[width=.22\textwidth]{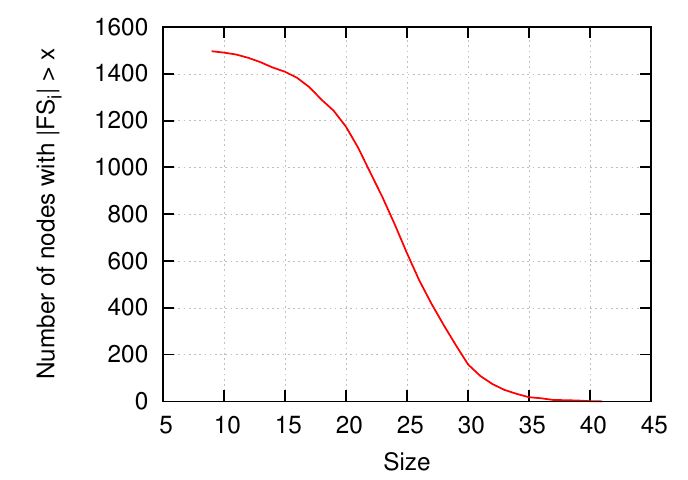}
        }
		\caption{Distribution of the sizes of the sets $\mathit{FS}_i$}
        \label{fig:fs_size}
    \end{center}
\end{figure}

Given these considerations, an intuitive choice is that of putting in $\mathit{FS}_i$  the set of nodes that $i$ is likely to encounter more often,
and in $\mathit{FI}_i(j)$ the expected inter contact times between $i$ and each of the nodes in $\mathit{FS}_i$.
To define $\mathit{FS}_i$ and $\mathit{FI}_i$ we again use the social behavior of node $i$ observed during the training period as an approximation of the its behaviour the rest of the trace.
To compute $\mathit{FS}_i$ set, we first consider the set of nodes $S_i$ that $i$ has seen at least once during the training period. 
Then, for every node $j \in S_i$ we calculate the number $d_{ij}$ of days in which $i$ met $j$, and compute the average value:
\[
\overline{d_i} = \frac{1}{|S_i|} \sum_{j \in S_i} d_{ij}
\]
Finally, we take as $\mathit{FS}_i$ the set of nodes $j$ such that $d_{ij} \geq \overline{d_i}$. In Figure \ref{fig:fs_size} we show the distribution of the size of the sets $\mathit{FS}_i$ in all the traces. For the Dartmouth
trace (Figure \ref{fig:fs_size-dartmouth}) the average size of $\mathit{FS}_i$ is $26.4$, for UCSD (Figure \ref{fig:fs_size-ucsd}) it is $9$, for Reality (Figure \ref{fig:fs_size-reality}) it is $12.3$ and for SWIM it is $24.3$. 
It is worth to observe that the sizes of the $\mathit{FS}_i$ sets don't change much from trace to trace despite the fact that in Dartmouth and SWIM the number of nodes is more than $20$ times bigger than
that of UCSD and Reality. We interpret this fact as an hint that our definition of $\mathit{FS}_i$ is meaningful and rather stable.

The values in $\mathit{FI}_i$ are computed starting from the inter-contacts observed between node $i$ and the nodes in $\mathit{FS}_i$ during the training period.
More in detail, for every pair of nodes~$(i, j)$ we computed $\mathit{FI}_i(j)$ as a the average number of days between contacts
of nodes~$i$ and~$j$.

\subsubsection{False Positives}

First, we tested the system in order to know if the honest nodes are always able to certificate their identity. We define a false positive as
the event that a legitimate user is not able to authenticate since he is temporarily out of his community for a long time.
To perform these tests for all the nodes we have set the value $k_i$ to $1$ and run the simulation as if no attack happens.
Setting $k_i$ to $1$ means that for a node it is sufficient to have just one valid signed  update to prove its identity. This permits
us to study the lower bound of the number of false positives the system can ensure given the definitions of $\mathit{FS}_i$
and $\mathit{FI}_i$ we provided above.

In the case of the Dartmouth trace, we have found that $43$ nodes (that is just the $4\%$ of all the $1101$ nodes) show false positives: 36 of them have only one false positive, while the rest have two false positives.
Moreover, the periods in which these nodes are not able to prove their identity always lasts one day at most. Considering the simple way in which we defined the sets $\mathit{FS}_i$ and the values $\mathit{FI}_i$ 
these results are pretty good. In the case of the UCSD trace our tests performed even better: only one node over the $32$ we tested (that is, the $3\%$ of the nodes) has false positives. In particular the node is not
able to prove its identity only for three time intervals, all of them lasting less than one day. We got a similar result in the Reality trace too: only one node over $45$ ($2\%$ of the nodes) had false positives 
and this happened only twice and for less than one day. Finally, in the SWIM trace no node has generated false positives. 

\subsubsection{Certificates Duration}

\begin{figure}[t]
    \begin{center}
        \subfigure[Dartmouth]
        {
            \label{fig:certificates-dartmouth}
            \includegraphics[width=.22\textwidth]{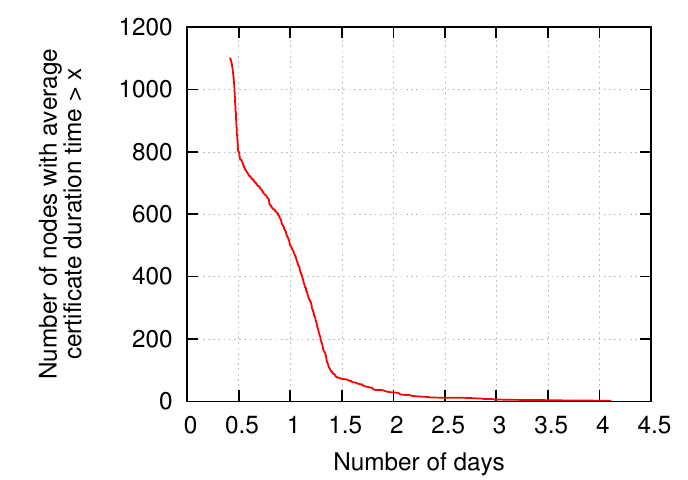}
        }
        \subfigure[UCSD]
        {
            \label{fig:certificates-ucsd}
            \includegraphics[width=.22\textwidth]{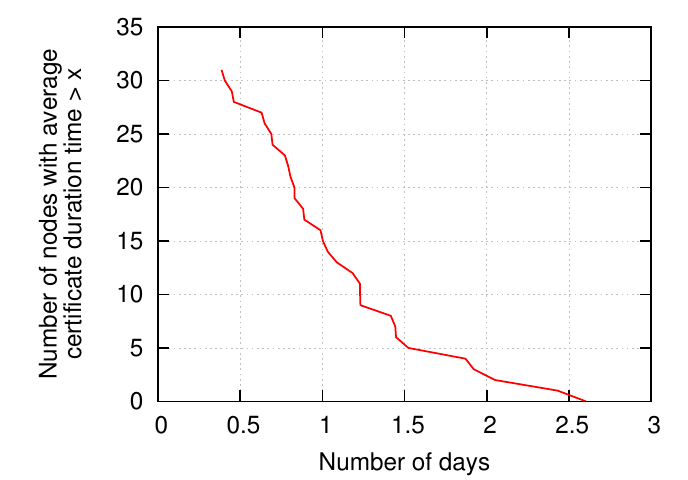}
        }
        \subfigure[Reality]
        {
            \label{fig:certificates-reality}
            \includegraphics[width=.22\textwidth]{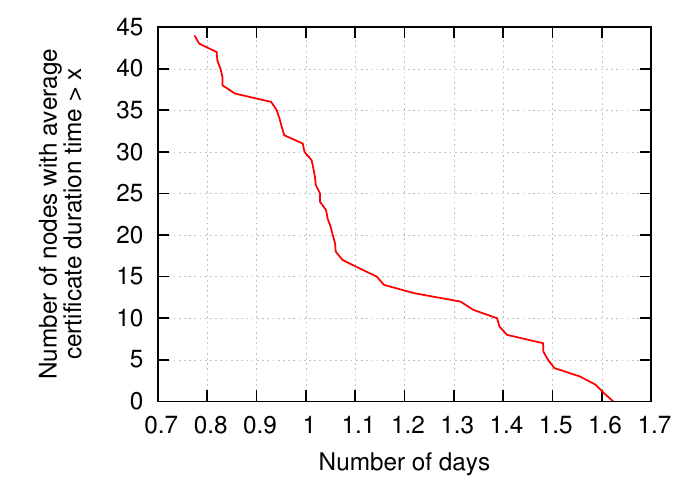}
        }
        \subfigure[SWIM]
        {
            \label{fig:certificates-swim}
            \includegraphics[width=.22\textwidth]{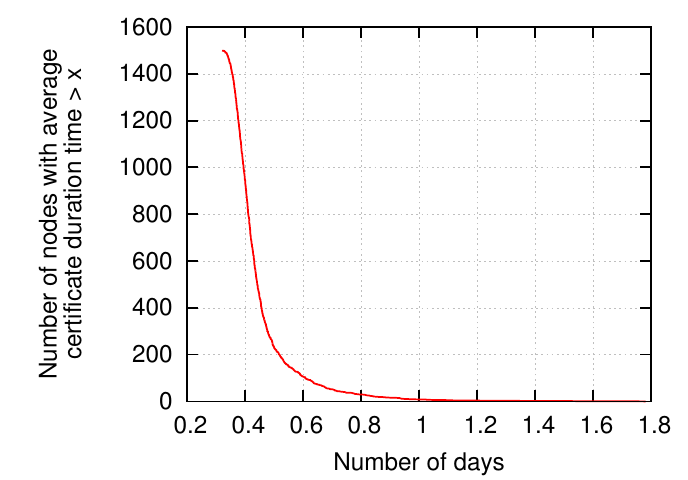}
        }
		\caption{Distribution of the average community certificates durations}
        \label{fig:certificates}
    \end{center}
\end{figure}

We have shown that the way we defined $\mathit{FS}_i$ and $\mathit{FI}_i$ is good enough to ensure that only a very small percentage of honest nodes suffers of false positives. Now it is time to study the effectiveness of the Community Certificates
sub-system in limiting the power of a clone attack by an outsider.
To do that, we study is the average time it takes for the community certificate of the nodes to expire in absence of contacts with nodes in the set
$\mathit{FS}_i$. This gives us a measure of 
the amount of time the adversary is able to use a cloned identity. The reason is that, once the adversary cloned an identity, he/she won't be able to use it inside the community of the victim or to renew it withouth being quickly detected by the Personal Marks sub-system. In such a situation, the only option left to the adversary is that of using the identity far from the victim's community.

When we studied the average duration of the community certificates, we used, for each node, the highest value $k_i$ for which the node is not affected by false positives. This gives us an idea of how much we can stress the
system to get the lowest possible average certificate duration given our definitions of $\mathit{FS}_i$ and $\mathit{FI}_i$. 
The results of our measurements are shown in Figure \ref{fig:certificates}. Each graph shows, for a given number of days $x$, what is the number of nodes whose certificate has, on average, a duration of more than $x$ days when cloned by an outsiders. So, the faster the better.
The first thing we can notice is that in all the traces the average duration of the certificate is never more than $2$ or $3$ days in the worst cases, and roughly between $0.5$ and $1.5$ in the majority of the cases. More in detail,
in the Dartmouth trace (Figure \ref{fig:certificates-dartmouth}) the community certificate has an average validity of less than one day and a half
for $93\%$ of the nodes, while only $3\%$ of the nodes have a certificate lasting, 
on average, more than two days. In the UCSD trace (Figure \ref{fig:certificates-ucsd}), the community certificates last on average at most one day and a half for $84\%$ of the nodes. This value is slightly smaller than the corresponding 
one found in the Dartmouth trace. However, in the worst case the average certificate duration in the UCSD trace is $2.6$ days as compared to the more than $3.5$ days of the Dartmouth trace. In the case of the Reality trace, the average 
certificate duration is at most one day and a half for the $91\%$ of the nodes and it is at most $1.62$ days long for all the nodes. This value is the minimum among the maximum average certificate duration we measured
in the three real traces we have used. Finally the simulated trace SWIM shows very good performance: $99\%$ of the nodes have an average certificate duration smaller than $1$ day and the maximum average certificate 
duration is $1.78$ days.

\begin{figure}[t]
    \begin{center}
        \subfigure[Dartmouth]
        {
            \label{fig:k-dartmouth}
            \includegraphics[width=.22\textwidth]{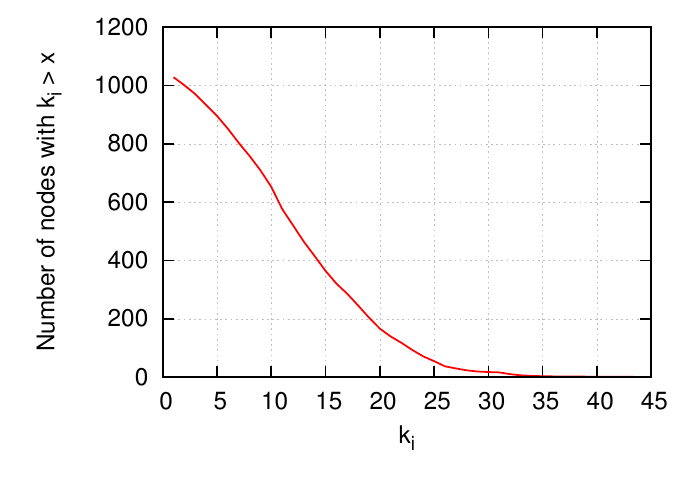}
        }
        \subfigure[UCSD]
        {
            \label{fig:k-ucsd}
            \includegraphics[width=.22\textwidth]{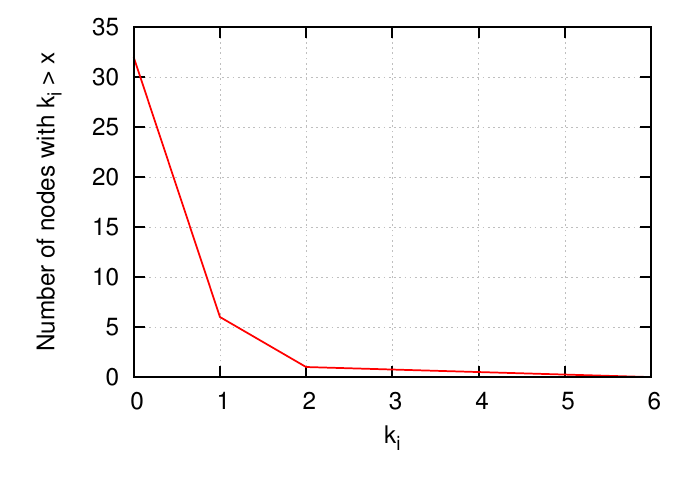}
        }
        \subfigure[Reality]
        {
            \label{fig:k-reality}
            \includegraphics[width=.22\textwidth]{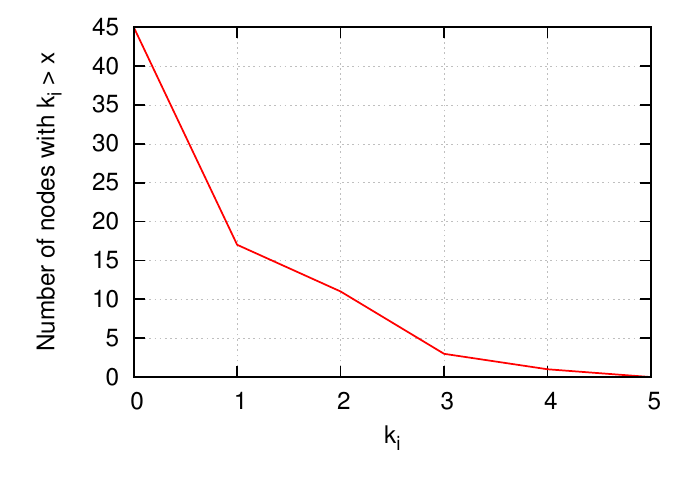}
        }
        \subfigure[SWIM]
        {
            \label{fig:k-swim}
            \includegraphics[width=.22\textwidth]{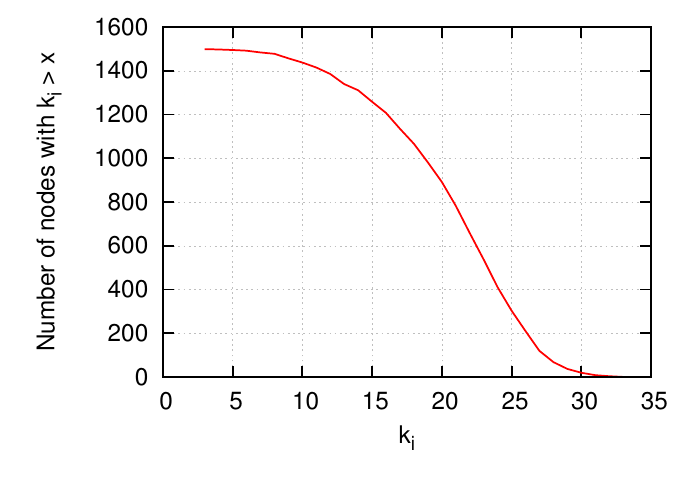}
        }
		\caption{Distribution of the values $k_i$}
        \label{fig:k}
    \end{center}
\end{figure}

Figure \ref{fig:k} shows the distributions of the values $k_i$ that we have found to be the maximum we could use on each node without
generating false positives. As we can see, there is a big difference between the optimal values
found for the Dartmouth and SWIM traces (Figures \ref{fig:k-dartmouth} and \ref{fig:k-swim}) and those found in the Reality and UCSD traces (Figures \ref{fig:k-reality} and \ref{fig:k-ucsd}). In the first two traces there is a big number
of nodes for which we have been able to select (automatically from the training period) high values of $k_i$ without producing false positives.
For instance, in the Dartmouth trace, for $93\%$ of the nodes we have been able to select a value $k_i$ greater than 1 and for $52\%$ of the nodes a value greater than $11$. In SWIM, the sets $\mathit{FS}_i$ are generally smaller than those found in Dartmouth (see Figure \ref{fig:fs_size}) and this translates into smaller values of $k_i$. Nevertheless in SWIM it is possible to
set a value $k_i$ greater than $1$ for $99\%$ of the nodes and a value greater than $10$ for $42\%$ of the nodes. On the other hand, in UCSD and Reality we can use a $k_i$ greater than one only for $19\%$ and for the $37.7\%$ of the nodes respectively.
This big difference can be explained by the fact that, as shown in Table \ref{tab:traces}, in Dartmouth and SWIM the nodes have a much higher number of contacts than in UCSD and Reality. In the case of Dartmouth this happens because 
the number of nodes is high and the trace has been collected using WiFi data rather than a short ranged bluetooth radio. On the other hand, in SWIM, since we are simulating a bluetooth trace, the nodes don't have a range as high as that of 
Dartmouth hence they tend to have contacts with smaller portions of the network. However, as already stressed in the previous sections, using the SWIM simulator we never miss a contact between two nodes and, moreover, we don't have data losses typical of the real traces.

Overall, we can say that the definitions we used  for $\mathit{FS}_i$ and $\mathit{FS}_j$ seem to work well in different scenarios. Indeed, they make it easy to the large majority of the honest nodes to keep
their community certificates up to date. At the same time, the adversary that clones an honest identity, will be able to use it for only short period of time that is in the order of one or two days at most.

\subsection{Using Fixed Infrastructure in the Community Certificates}
\label{sec:celltowers}

So far we have assumed that each node receives its signed timestamps from the other nodes in its community.
However, the Community Certificates protocol is actually more general than that. Indeed, the certificate authority may potentially put 
into the set $\mathit{FS}_i$ the ID of any kind of device with a private/public key pair that is able to send signed timestamps
to the nodes that enter its range of communication. For instance, we could think of a WiFi access point in a university campus.
In such a scenario it would be perfectly reasonable for a student to have stored in the set $\mathit{FS}_i$ of his/her smartphone the public key of the access points of the university buildings he/she frequents the most.
There would be three good reasons to do that. The first one is that there are locations, just like friends, that we ``meet'' more often
than others; it is easy to check experimentally that the pattern of meetings is similar to that of two nodes carried by humans.
The second one is that the access points have fixed locations and are, usually, always on. The third one is that it is supposedly 
harder to tamper with an access point and to move it to another location rather than a mobile device. This would make it more difficult
for the adversary to refresh the community certificate of a cloned identity $i$ by compromising a given number of nodes in
the set $\mathit{FS}_i$ (see Section \ref{subsection:multiple_attacks}).

\begin{figure}[t]
    \begin{center}
        \subfigure[Sizes of the sets $\mathit{FS}_i$]
       {
            \label{fig:dartmouth.APS-fs_size}
            \includegraphics[width=.22\textwidth]{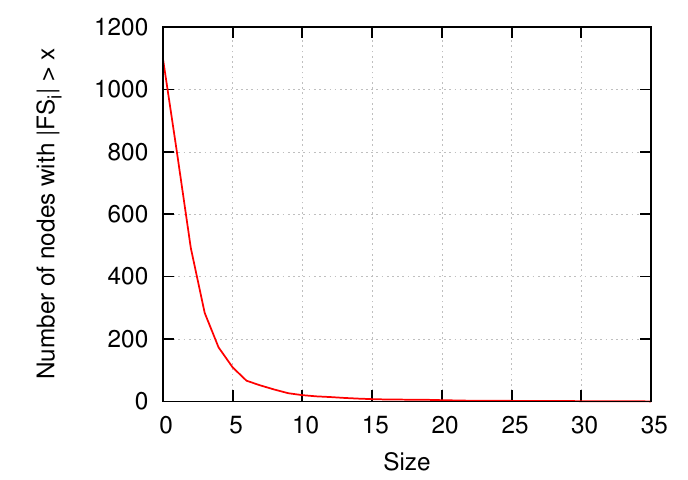}
        }
        \subfigure[Average certificate durations]
        {
            \label{fig:dartmouth.APS-certificates}
            \includegraphics[width=.22\textwidth]{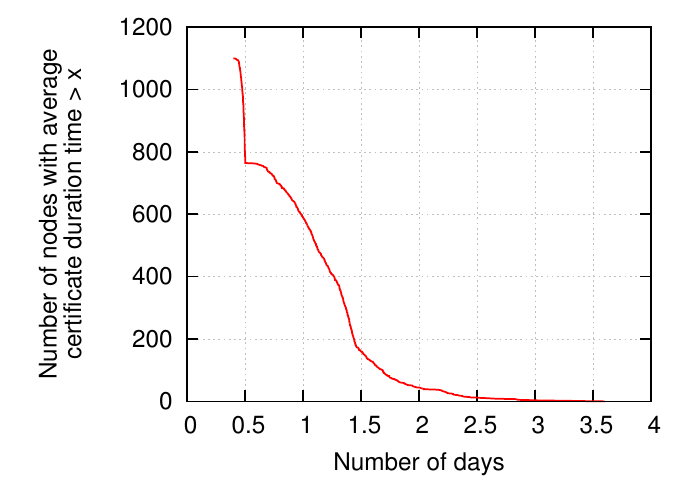}
        }
        \subfigure[Values $k_i$]
        {
            \label{fig:dartmouth.APS-k}
            \includegraphics[width=.22\textwidth]{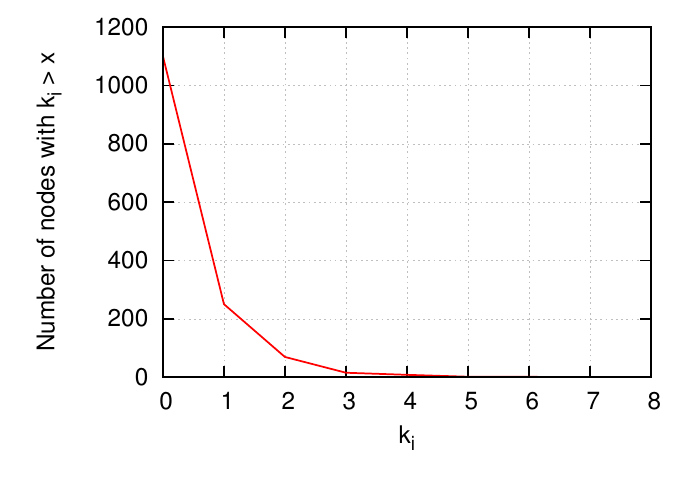}
        }
		\caption{Dartmouth (Using Access Points)}
        \label{fig:dartmouth.APS}
    \end{center}
\end{figure}

\begin{figure}[t]
    \begin{center}
        \subfigure[Sizes of the sets $\mathit{FS}_i$]
       {
            \label{fig:ucsd.APS-fs_size}
            \includegraphics[width=.22\textwidth]{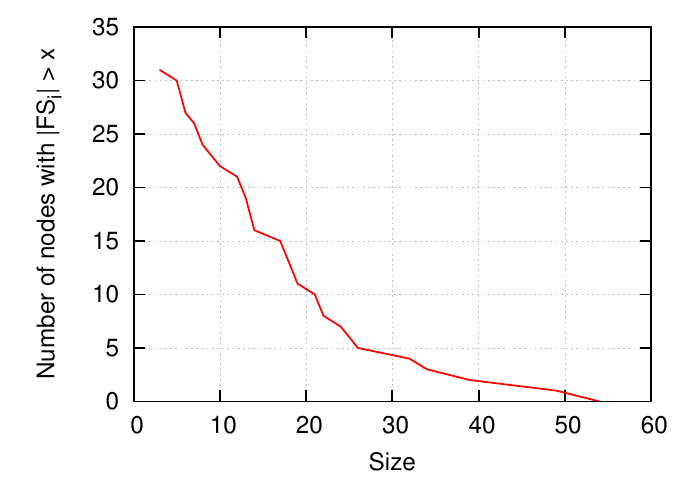}
        }
        \subfigure[Average certificate durations]
        {
            \label{fig:ucsd.APS-certificates}
            \includegraphics[width=.22\textwidth]{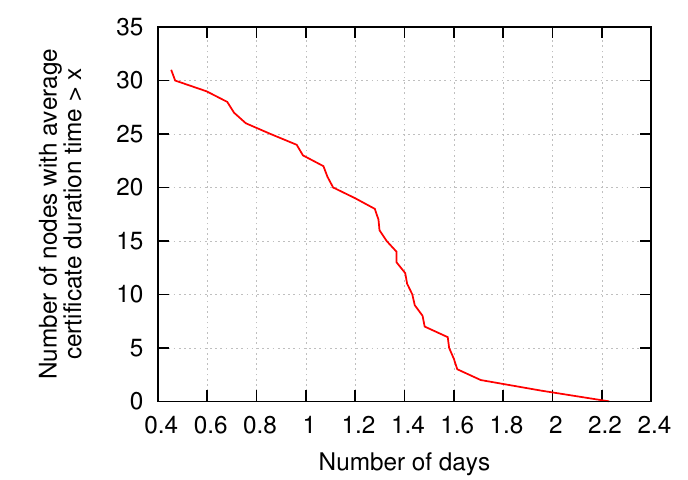}
        }
        \subfigure[Values $k_i$]
        {
            \label{fig:ucsd.APS-k}
            \includegraphics[width=.22\textwidth]{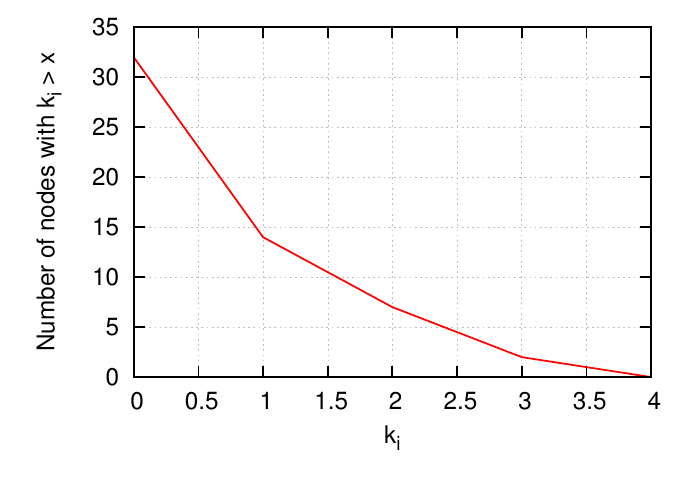}
        }
		\caption{UCSD (Using Access Points)}
        \label{fig:ucsd.APS}
    \end{center}
\end{figure}

For these reasons we found it interesting to test the performance of Community Certificates in a situation where the certificate authority
puts in $\mathit{FS_i}$ the access points of the most frequented locations of each node. To do that, we used the Dartmouth and UCSD traces since they have been collected exactly recording the associations between mobile nodes and the access point of the 
Dartmouth and University of California San Diego campuses respectively (see Section~\ref{subsection:datasets}). 
We performed the tests with the same definitions of $\mathit{FS_i}$ and $\mathit{FI_i}$ we used in Section~\ref{subsec:commcert}, though,
this time, we are interested in the contacts between nodes and access points rather than between pair of nodes.
The results are shown in Figures~\ref{fig:dartmouth.APS} and \ref{fig:ucsd.APS} . 

The first difference we can notice between these results and the previous ones is the change in the sizes of the sets $\mathit{FS_i}$. In the case of the Dartmouth trace (Figure \ref{fig:dartmouth.APS-fs_size}) 
the average size of $\mathit{FS_i}$ is now only $3$, while before it was $26$.  On the contrary, in the UCSD (Figure \ref{fig:ucsd.APS-fs_size}) trace the size of the $\mathit{FS_i}$
increased from an average of $9$ to an average of $18$. This is a consequence of the fact that in Dartmouth the contacts between the nodes happened in a restricted set of places, while, in UCSD, 
the nodes were generally more mobile. Note that these values are small as compared to the total number of access points, that is about $500$ in both traces. The second difference is that the average certificate duration of the nodes slightly increases with respect to our previous results
(remember, however, that in these experiments we have removed the mobile friends!).
In the case of Dartmouth (Figure~\ref{fig:dartmouth.APS-certificates}), the percentage of nodes whose community certificate has an average validity of less than one day has dropped from $54\%$ to $46\%$. Also, the community certificates have an average duration of 
less than $1.5$ days for $85\%$ of the nodes while before it was for $93\%$. In the case of UCSD (Figure \ref{fig:ucsd.APS-certificates}) $31\%$ of the nodes have a community certificate lasting 
on average less than one day (before it was $53\%$) while for $81\%$ it lasts less than one day and a half (was $84\%$).
To summarize, the results are not as good as with real mobile friends. However, they are good enough to make it reasonable
to complement the set~$\mathit{FS_i}$ with a small number of entities that are part of the fixed infrastructure. This is not going
to reduce the performance considerably and helps build a stronger system against coordinate attacks to multiple mobile nodes.
Lastly, Figures \ref{fig:dartmouth.APS-certificates} and \ref{fig:ucsd.APS-certificates} show distribution of the values of $k_i$.

\section{Add-ons and Caveats}
\label{sec:caveats}

Community Certificates can be complemented and extended in such a way to design
more sophisticated and complete versions. Here we discuss a few of these add-ons.

\paragraph{Travelling}
If we travel, especially alone, our community certificate is going to expire soon.
While this is generally true, we can easily design a few solutions:
(i) The user can suspend the community checks on the certificate during the travel; (ii) if we often travel between two cities,
like Rome and San Diego, the user can prepare two different \emph{profiles}, in the same certificate, that can
be selected when changing community. Of course, it is fundamental that these solutions can be used
only by the the legitimate owner by performing the burdensome alternative authentication that is
used in case of exceptional events.
\paragraph{Complex patterns}
It is fairly easy to extend Community Certificates in such a way that it is not enough to get $k$ fresh signed
timestamps
from your circle of friends. Rather, the certificate needs $k_1$ fresh signed timestamps from
one sub-community (for example your family) and $k_2$ fresh signed timestamps from another
sub-community (for example your colleagues) to be valid. A community certificate like this can easily be more efficient
and more useful in particular situations. For example when we need at least one signature from one
of our bosses to be able to perform critical operations. Or at least one signature from a fixed
infrastructure in the building of our company, for example.
\paragraph{Electronic handshakes}
In addition to standard physical contacts, Community Certificates can be extended to use \emph{electronic
handshakes}. Assume that, when meeting a friend, you use a particular secure "handshake" made with
your devices (like bumping the two together, for example). Handshakes can be used at the place of physical
proximity. In this way, contacts are more trustworthy, but, on the other hand, they need human intervention
and are much fewer in number. Alternatively, handshakes could be used to exchange a stronger
signed timestamp. By using \emph{complex patterns}, one or more of these handshakes can be required
to enable a higher level of authentication to perform critical operations.

\section{Conclusions}
\label{sec:conclusions}

In this paper we introduce Personal Marks and Community Certificates. The fundamental idea of Community
Certificates is that, in networks
of mobile people, authentication can be based on the notion of community, and nodes can authenticate by
showing that they indeed meet the people that are part of their community. While the social structure of these
networks has been extensively used in networking, to the best of our knowledge this is the first time that
this has been used as a biometric to authenticate device. We also present
Personal Marks, a way a community can use to protect itself against insiders performing a clone attack. The
combined used of these mechanisms deliver an excellent protection of the social mobile network against
the clone attack. Indeed our experiments show that the detection is always correct, and fast enough
considered the slow dynamics of the trace we have used. In any real system, we believe that the much faster
dynamics of meetings can help deliver much faster detection times.

\bibliographystyle{abbrv}

\end{document}